\documentclass[11pt]{article}

\usepackage[utf8]{inputenc}
\usepackage[T1]{fontenc}
\usepackage[margin=1in]{geometry}
\usepackage{abstract}
\usepackage{titlesec}
\usepackage{enumitem}
\usepackage{hyperref}
\usepackage{xcolor}
\usepackage{listings}

\hypersetup{
  colorlinks=true,
  linkcolor=black,
  citecolor=black,
  urlcolor=blue,
  pdftitle={Kutti AI: A Voice-First, Offline-Capable Learning Companion for Visually-Impaired Children},
  pdfauthor={Kadharmoideen Fadurudeen}
}

\titlespacing*{\section}{0pt}{1.4ex plus 1ex minus .2ex}{1ex plus .2ex}
\titlespacing*{\subsection}{0pt}{1.1ex plus 1ex minus .2ex}{0.8ex plus .2ex}

\title{\textbf{Kutti AI: A Voice-First, Offline-Capable Learning Companion with Real-Time Struggle Detection for Visually-Impaired Children}}

\author{
  Kadharmoideen Fadurudeen\\
  Independent Researcher\\
  \texttt{kadhar.mfn@gmail.com} \quad \url{https://kadhar.dev}
}
\date{}

\begin{document}
\maketitle

\begin{abstract}
\noindent
Most educational technology for children is built around visual interfaces, which excludes the many children worldwide who live with visual impairment---an estimated 1.4 million children are blind and many more have low vision. We present Kutti AI, a voice-first learning companion designed so that audio is the primary and sufficient interface: children learn curriculum concepts through spoken conversation, respond by speaking, and receive spoken feedback, with no reliance on visual elements. The system contributes three practical mechanisms for accessible, adaptive learning on commodity mobile hardware: (1) a multi-signal \emph{struggle-detection} engine that combines response-latency analysis, wrong-attempt tracking, and keyword-based hesitation detection to decide, in real time, when to offer hints or simplify a question; (2) a multi-layered \emph{cross-language answer-matching} pipeline that combines language-aware translation/transliteration, Levenshtein-based fuzzy matching, and text normalization so that children are not penalized for code-switching or pronunciation variation; and (3) an \emph{offline-first} speech pipeline using an on-device automatic speech recognition (ASR) model, enabling use in low-connectivity settings common in underserved communities. We describe the architecture, the interaction flow, and the design decisions that prioritize accessibility, and we report qualitative observations from a hackathon prototype supporting English and Tamil. We discuss lessons learned and outline a path toward formal evaluation with target users. Kutti AI illustrates how a small, carefully-engineered voice-first system can lower both accessibility and financial barriers to early education.

\vspace{0.6em}
\noindent\textbf{Keywords:} accessibility, assistive technology, voice user interfaces, adaptive learning, speech recognition, educational technology, visual impairment, offline AI
\end{abstract}

\section{Introduction}

For most children, early learning is mediated visually---recognizing letters, following a screen, tapping images. This assumption is so deeply embedded in educational software that it quietly excludes children who cannot rely on sight. Global health estimates indicate that roughly 1.4 million children under age 15 are blind, with many millions more affected by visual impairment more broadly, and many of them lack access to quality educational technology precisely because that technology presumes a visual user.

Two barriers compound the problem. The first is \emph{modality}: apps that require reading text, seeing images, or navigating a visual layout are difficult or impossible to use without sight. The second is \emph{economic}: much assistive and educational software sits behind subscriptions or hardware requirements that place it out of reach for families in underserved communities, where the need is often greatest.

This paper presents Kutti AI, a learning companion that addresses both barriers. Kutti AI is designed to be \emph{voice-first in the strong sense}: the entire learning loop---presenting a concept, posing a question, capturing the answer, detecting difficulty, and giving feedback---is conducted through speech, and the application is intended to be distributed free of charge. Rather than treating voice as an add-on to a visual app, we treat audio as the primary interface and design the interaction, the data model, and the feedback loop around it.

The central engineering question is not whether a device can convert speech to text, but how a system should \emph{behave} once it is listening: how to tell when a child is struggling and respond supportively, how to fairly judge spoken answers from children who mix languages and mispronounce words, and how to do all of this on ordinary phones that may lack a reliable internet connection. Our contributions are:

\begin{enumerate}[leftmargin=1.4em]
  \item A \textbf{multi-signal struggle-detection engine} that fuses temporal (pause), behavioral (wrong-attempt), and linguistic (hesitation-keyword) signals to trigger real-time hints and difficulty adjustment.
  \item A \textbf{cross-language answer-matching pipeline} combining language-aware translation/transliteration, fuzzy string matching, and normalization, enabling tolerant evaluation of young children's spoken responses.
  \item An \textbf{offline-first architecture} using on-device ASR and locally-cached lessons, so the core experience works without connectivity, with progress synchronization when a network is available.
\end{enumerate}

We describe the system, report qualitative prototype observations from the Half Baked hackathon (supporting English and Tamil), discuss design trade-offs and lessons, and outline a plan for evaluation with visually-impaired learners.

\section{Related Work}

Kutti AI draws on three strands of prior work: assistive technology for visually-impaired users, voice-based education (particularly for visually-impaired children), and the well-documented difficulty of recognizing children's speech.

\paragraph{Assistive and voice-based technology for visually-impaired users.}
Assistive technology for visually-impaired users is a mature area, spanning scene-description tools, text-narration applications, and screen readers that make conventional software navigable without sight. Voice interfaces are a natural fit for this population, and a broad literature survey of accessibility research at CHI and ASSETS documents the field's growth and its recurring themes~\cite{mack2021}. Our work is squarely within this tradition but focuses on a learning application rather than navigation or general access.

\paragraph{Voice-based education for visually-impaired children.}
The closest prior work to Kutti AI is SEEDS, a voice-based education platform for children and teachers with visual impairments, which combines a teacher application with an interactive voice response (IVR) system to deliver educational content, and which is motivated by the same context---the large population of children with visual impairments in India and their limited access to accessible resources~\cite{poddar2024}. Related efforts have explored voice-driven learning tools co-designed with educators and visually-impaired children, for example accessible programming environments operated by voice~\cite{pires2020}. Kutti AI differs from these systems in three respects that motivate our contributions: it runs its speech recognition \emph{on-device and offline} rather than through a network/IVR dependency; it adds \emph{real-time, multi-signal struggle detection} to adapt difficulty automatically; and it performs \emph{tolerant cross-language answer matching} so that young children are not penalized for code-switching or mispronunciation.

\paragraph{Young children and voice interfaces.}
Foundational work on how young children interact with voice input systems highlights both the promise and the pitfalls of speech as a children's interface~\cite{lovato2015}, informing our decision to make the interaction forgiving and to provide immediate spoken support.

\paragraph{The difficulty of children's speech recognition.}
A central design assumption of Kutti AI---that spoken answers from children must be matched \emph{tolerantly}---is grounded in a large body of evidence that automatic speech recognition (ASR) for children is substantially less accurate than for adults, owing to higher acoustic and pronunciation variability, disfluencies, and developmental differences in the vocal tract~\cite{potamianos2003}. A recent systematic review confirms these challenges persist across modern systems~\cite{bhardwaj2022}, and work adapting the Whisper model specifically to child speech shows both the difficulty and the active interest in this problem~\cite{jain2023}. Because Kutti AI uses Whisper for on-device recognition~\cite{radford2022}, these findings directly justify the fuzzy, normalization-based, and cross-language matching layers described in Section~\ref{sec:matching}, which are built on classic edit-distance techniques~\cite{levenshtein1966}.

In short, Kutti AI occupies the intersection of these strands: a voice-first, offline-capable learning system for visually-impaired children whose adaptive-tutoring logic is engineered around the known realities of children's speech.

\section{System Overview and Requirements}

We derived four requirements from the target users and deployment context:

\begin{itemize}[leftmargin=1.4em]
  \item \textbf{R1 --- Audio sufficiency.} Every function must be reachable and completable through speech and audio feedback alone; visual elements may exist for sighted caregivers but must never be \emph{required}.
  \item \textbf{R2 --- Supportive adaptivity.} The system must recognize when a child is struggling and respond with encouragement, hints, or simplified questions, so the child is not left stuck or discouraged.
  \item \textbf{R3 --- Tolerant answer evaluation.} Because young children mispronounce words and mix languages, correctness judgments must be robust to pronunciation variation and code-switching.
  \item \textbf{R4 --- Offline capability.} The core learning experience must function without internet access, with graceful synchronization when connectivity returns.
\end{itemize}

\subsection{Technology Stack}
Kutti AI is a cross-platform mobile application built with React Native (Expo) and TypeScript. Speech recognition uses an on-device port of the Whisper ASR model, enabling offline transcription. Speech synthesis uses the platform text-to-speech engine for multilingual lesson delivery, and audio capture is handled through the platform audio-recording library. Persistent data---lessons, session records, and aggregate progress---is stored in a PostgreSQL backend (Supabase), with row-level security enforcing per-user data isolation and real-time features reserved for future collaborative capabilities. Lessons are cached locally so that transcription and lesson delivery do not depend on the network.

\section{Interaction Design}

The learning loop is designed as a spoken conversation with five stages:

\begin{enumerate}[leftmargin=1.4em]
  \item \textbf{Welcome.} On launch, the app speaks a welcome message in the selected language. A single large, haptic-enabled control lets the child begin without needing to locate small visual targets.
  \item \textbf{Lesson delivery.} Lesson steps are spoken via text-to-speech, with each question read clearly and paced with natural pauses so the child can process it.
  \item \textbf{Voice response.} The child speaks an answer, captured either by tapping a large microphone control or automatically in a hands-free mode, and transcribed on-device.
  \item \textbf{Adaptive response.} The system evaluates the answer and the surrounding behavioral signals and responds: encouragement for a correct answer, a hint when it detects hesitation, or a simplified question after repeated difficulty (Section~\ref{sec:struggle}).
  \item \textbf{Summary.} On completion the child hears an engagement score (0--100) and encouraging feedback, and the session is persisted for progress tracking.
\end{enumerate}

To satisfy R1 beyond the core loop, the interface is screen-reader compatible with labeled controls, uses large touch targets, provides haptic confirmation for actions, and offers a fully audio-only mode. For low-vision (rather than no-vision) users and sighted caregivers, a high-contrast color scheme is used; these visual affordances are strictly supplementary.

\section{Real-Time Struggle Detection}
\label{sec:struggle}

The core of Kutti AI's adaptivity (R2) is a struggle-detection engine that fuses three independent signals and acts the moment any of them indicates difficulty.

\paragraph{Signal 1 --- Response latency.}
The system measures the elapsed time between the end of a spoken question and the beginning of the child's response. A pause beyond a threshold (2.5 seconds in the prototype) is treated as a sign of difficulty and triggers a spoken hint. This threshold was chosen empirically to balance responsiveness against giving a child enough time to think; it is a natural candidate for per-child calibration in future work.

\paragraph{Signal 2 --- Wrong-attempt tracking.}
The system counts incorrect attempts on a given question. After two incorrect attempts it presents a \emph{simplified} version of the question---authored in advance as part of the lesson content---so the child can experience success rather than repeated failure.

\paragraph{Signal 3 --- Hesitation keywords.}
The transcribed response is scanned for hesitation cues such as ``don't know,'' ``not sure,'' or ``help.'' When such a cue is detected, the system provides encouragement and additional support rather than marking the answer wrong.

Because the three signals are evaluated together and any one can fire, the system reacts to several different manifestations of difficulty---silence, error, and verbalized uncertainty---using only lightweight, on-device computation, which keeps the loop fast and offline-capable. Conceptually:

\begin{lstlisting}
onChildResponse(transcript, timing, attempts):
    if timing.sincePrompt > PAUSE_THRESHOLD:  offerHint()
    if attempts.wrong >= WRONG_ATTEMPT_LIMIT:  showSimplifiedQuestion()
    if containsHesitation(transcript):         provideEncouragement()
\end{lstlisting}

\section{Cross-Language Answer Matching}
\label{sec:matching}

Meeting R3---fairly judging young children's spoken answers---is the subtlest technical problem in the system. A child may answer ``one'' when the expected answer is the Tamil equivalent, or produce a near-miss pronunciation. Naive string equality would wrongly mark such answers incorrect and discourage the learner. Kutti AI therefore evaluates answers through a three-layer pipeline.

\paragraph{Layer 1 --- Language-aware matching.}
The system detects when the child has responded in a different language than the lesson expects and reconciles the two by translating or transliterating before comparison, so an English answer to a Tamil prompt (or vice versa) can still match.

\paragraph{Layer 2 --- Fuzzy matching.}
To absorb pronunciation variation and minor transcription errors, the system computes similarity from Levenshtein edit distance and accepts responses above a similarity threshold (0.6 in the prototype), rather than requiring exact equality.

\paragraph{Layer 3 --- Normalization.}
Before comparison, transcripts are lowercased, stripped of leading filler words (``um,'' ``uh,'' ``hmm''), cleared of punctuation, and whitespace-normalized, so superficial differences do not affect the judgment.

Applied in sequence, these layers let the system accept a wide range of legitimate answers---cross-language, mispronounced, or disfluent---while still distinguishing genuinely incorrect responses. This tolerance is not a mere convenience: for a child who cannot see a spelled answer, being fairly understood when speaking is central to the experience being usable at all.

\section{Data Model}

The backend schema is intentionally small and analytics-friendly. A \textbf{Lessons} table stores curriculum content, holding each lesson's steps---questions, expected answers, hints, and simplified variants---as structured JSON, together with metadata such as category, difficulty, language, and region. A \textbf{Learning Sessions} table records each attempt, including the engagement score, count of correct answers, and number of struggle moments, plus a structured session record. A \textbf{User Progress} table aggregates history---total lessons, average score, and completion counts---to support progress tracking and future analytics. Row-level security isolates each user's data.

\section{Offline-First Design}

Reliable connectivity cannot be assumed in many of the communities Kutti AI aims to serve, motivating R4. Two decisions make the core experience network-independent: speech recognition runs on-device via an embedded Whisper model, and lesson content is cached locally. As a result, transcription, lesson delivery, struggle detection, and answer matching all operate without a network. Progress synchronizes to the backend opportunistically when connectivity is available. This design trades some on-device compute and storage for availability---the right trade-off when the alternative is that the child cannot use the app at all.

\section{Prototype and Qualitative Observations}

Kutti AI was built as a working prototype during the Half Baked hackathon, whose challenge was to build a startup idea addressing a real problem. The prototype supports English and Tamil, with a language architecture designed so that additional languages can be added by supplying localized content and feedback rather than changing the core logic. Informal response to the prototype was positive, and the exercise validated the central interaction loop end-to-end: spoken lesson delivery, on-device transcription, multi-signal struggle detection, tolerant answer matching, and spoken adaptive feedback.

We are careful not to overstate these results. They are qualitative and formative; the prototype has not yet been evaluated with visually-impaired children under controlled conditions. What the prototype establishes is \emph{feasibility}---that a fully voice-first, offline-capable, adaptive learning loop can be implemented on commodity mobile hardware---and a concrete platform on which to base a proper user study (Section~\ref{sec:future}).

\section{Lessons Learned}

Several lessons emerged that may help others building voice-first learning tools:

\begin{itemize}[leftmargin=1.4em]
  \item \textbf{Design for being \emph{misheard}, not just \emph{heard}.} For young children, ASR errors and disfluencies are the norm, not the exception. The tolerant, multi-layer answer-matching pipeline was not an optional refinement; without it, correct answers were frequently rejected and the experience felt punitive. Robust matching is a first-class requirement in a children's voice interface.
  \item \textbf{Struggle is multi-modal, so detection should be too.} No single signal reliably captures difficulty. Silence, wrong answers, and verbalized uncertainty each miss cases the others catch; fusing them produced far more humane behavior than any one alone.
  \item \textbf{Offline capability changes who the system can reach.} Moving ASR on-device was the decision that made the app relevant to low-connectivity settings. Accessibility and connectivity constraints are intertwined for this population.
  \item \textbf{Simplification must be authored, not improvised.} Pre-authoring simplified variants of each question, rather than generating them on the fly, kept behavior predictable and appropriate for young learners in the prototype---a useful constraint under hackathon conditions and for offline operation.
\end{itemize}

\section{Limitations and Future Work}
\label{sec:future}

The primary limitation is evaluation: we report feasibility and qualitative impressions, not measured learning or usability outcomes with target users. Our priority is a study with visually-impaired children and their educators, measuring task completion, engagement, and learning gains, and validating the struggle-detection thresholds against observed behavior. Fixed parameters---the 2.5-second pause threshold, the two-attempt limit, the 0.6 similarity threshold---should become adaptive and personalized. Planned functional extensions include additional languages beyond English and Tamil, a caregiver dashboard for progress, optional gamification, and authoring tools so educators can create curriculum-aligned lessons. Finally, keyword-based hesitation detection could be extended toward richer prosodic or acoustic sentiment analysis while preserving on-device, privacy-respecting operation.

\section{Conclusion}

Kutti AI shows that the exclusion of visually-impaired children from educational technology is an engineering choice, not an inevitability. By making audio the sole required modality, detecting struggle from multiple real-time signals, judging spoken answers tolerantly across languages, and running its core loop offline on ordinary phones, the system lowers both the accessibility and the economic barriers to early learning. The hackathon prototype demonstrates feasibility for English and Tamil; the next step is rigorous evaluation with the children the system is meant to serve. We hope the design patterns described here---audio sufficiency, multi-signal struggle detection, tolerant cross-language matching, and offline-first operation---are useful to others building inclusive learning technology.

\section*{Acknowledgments}
Kutti AI was created during the Half Baked hackathon. The author thanks the organizers and participants for feedback on the prototype.

\end{document}